\documentclass[12pt]{article}
\usepackage{graphicx}
\usepackage{amssymb}
\usepackage{cite}

\newcommand{\goes}{\rightarrow}                          %
\newcommand{\GeV}{\; \mathrm{GeV}}                       %
\newcommand{\TeV}{\; \mathrm{TeV}}                       %
\newcommand{\beq}{\begin{equation}}                      %
\newcommand{\eeq}{\end{equation}}                        %
\newcommand{\bea}{\begin{eqnarray}}                      %
\newcommand{\eea}{\end{eqnarray}}                        %
\newcommand{\lsp}{\tilde{\chi}}                          %
\newcommand{\mlsp}{m_{\lsp}}                             %
\newcommand{\relic}{\Omega_{\lsp}\,h_0^2}                %
\newcommand{\sxsec}{\sigma_{scalar}}                     %
\newcommand{\almuon}{\alpha_{\mu}^{\mathrm{SUSY}}}       %
\newcommand{\etal}{\textit{et. al.}}                     %

\newsavebox{\sboxpubnumber}
\newsavebox{\sboxpubdate}

\newcommand{\pubnumber}[1]{\begin{lrbox}{\sboxpubnumber}{\begin{tabular}{l} #1 \\
                                 \usebox{\sboxpubdate}
                                 \end{tabular}}
                           \end{lrbox}
                           \pubblock}
\newcommand{\Title}[1]{\begin{center} {\Large #1 } \end{center}}
\newcommand{\Author}[1]{\begin{center}{ \sc #1} \end{center}}
\newcommand{\Address}[1]{\begin{center}{ \it #1} \end{center}}

\newcommand{\pubblock}{\rightline{
                        \usebox{\sboxpubnumber}}}
\newenvironment{Abstract}{\begin{quotation}  }{\end{quotation}}
\newenvironment{Presented}{\begin{quotation} \begin{center}
             PRESENTED AT\end{center}\bigskip
      \begin{center}\begin{large}}{\end{large}\end{center}
      \end{quotation}}
\newcommand{\Acknowledgements}{\bigskip  \bigskip \begin{center} \begin{large}
             \bf ACKNOWLEDGEMENTS \end{large}\end{center}}

\begin{document}
\def\thefootnote{\fnsymbol{footnote}}
\setcounter{footnote}{0}
\begin{titlepage}
%
\pubnumber{ hep-ph/0112134\\
            HEPHY-PUB 750/01\\
            CTP-TAMU 32/01, ACT-10/01  } 

\vfill
\Title{Combining Supersymmetric Dark Matter with
Recent  Accelerator Data\footnote{PRESENTED by V.~C.~Spanos}}
\vfill
\Author{A.~B.~Lahanas} 
\Address{University of Athens, Physics Department,  
Nuclear and Particle Physics Section,\\  
GR--15771  Athens, Greece}
\vfill
\Author{D.~V.~Nanopoulos}
\Address{Department of Physics,  
         Texas A \& M University, College Station,  
         TX~77843-4242, USA, 
         Astroparticle Physics Group, Houston 
         Advanced Research Center (HARC), Mitchell Campus, 
         Woodlands, TX 77381, USA, and \\ 
         Academy of Athens,  
         Chair of Theoretical Physics,  
         Division of Natural Sciences, 28~Panepistimiou Avenue,  
         Athens 10679, Greece}
\vfill
\Author{V.~C.~Spanos}
\Address{Institut f\"ur Hochenergiephysik der \"Osterreichischen Akademie
der Wissenschaften,\\
A--1050 Vienna, Austria}
\vfill
\begin{Abstract}
In the framework of the Constrained Minimal Supersymmetric Standard Model 
we discuss the impact of  the recent experimental
information, especially from the E821 Brookhaven experiment  
on  $g_{\mu}-2$ along with the light Higgs boson mass bound
from LEP,  in delineating regions of the
parameters which are consistent with cosmological data. The effect
of these to the Dark Matter direct searches is also discussed. 

\end{Abstract}
\vfill
\begin{Presented}
    COSMO-01 \\
    Rovaniemi, Finland, \\
    August 29 -- September 4, 2001
\end{Presented}
\vfill
\end{titlepage}

\section{Introduction}
Supersymmetry (SUSY)  is a landmark
in our efforts to construct a unified theory of all
fundamental interactions observed in nature. At very high energies, close
to the Planck scale ($M_P$) it is indispensable  in 
constructing consistent string
theories, and at low energies ($\sim 1 \TeV$)
it seems unavoidable if the gauge hierarchy problem is to
be resolved. Such a resolution provides a measure of the supersymmetry
breaking scale $M_{SUSY} \thickapprox \mathcal{O}(1 \TeV) $. 
There is  indirect evidence for such a
 low-energy supersymmetry breaking scale, from the unification
of the gauge couplings \cite{Kelley} and from  lightness
of the Higgs boson as determined from precise electroweak measurements,
mainly at LEP \cite{EW}. Furthermore, such a low energy SUSY breaking
scale is also favored cosmologically. As is well known, $R$-parity
conserving  SUSY models, contain in the sparticle spectrum a stable,
neutral particle, identifiable with the lightest neutralino ($\lsp$),
referred to 
as the LSP \cite{Hagelin}.
It is important \cite{Hagelin} that
such a LSP with mass, as low-energy SUSY entails, in the
 $ 100\GeV - 1\TeV $ region, may indeed provide the right form and amount
of the highly desirable astrophysically and cosmologically Dark Matter (DM).
The latest data about Cosmic
Microwave Background (CMB) radiation 
anisotropies \cite{cmb} not only favour
a flat ($k=0$ or $\Omega_0=1$),
inflationary Universe, but they  also determine a matter density
$\Omega_M h_0^2 \thickapprox 0.15 \pm 0.05$.
Taking into account the simultaneously determined baryon density
$\Omega_B h_0^2 \thickapprox 0.02$, and the rather tiny
neutrino density, they result to
\beq
\Omega_{DM} h_0^2 = 0.13 \pm 0.05 \,.
\label{cbound}
\eeq

If we assume that all DM is supersymmetric due to LSP,
i.e. $\Omega_{DM} \equiv \Omega_{\lsp}$,
it is tempting to combine the bound of Eq.\ref{cbound} with
other  presently available constraints from particle physics,
such as the lower bound on the
mass of the Higgs bosons ($m_{h} \geq 113.5 \GeV$) provided by LEP \cite{LEP}
and the recent results from the BNL E821 experiment \cite{E821} on the 
anomalous magnetic moment of the muon 
($\delta \alpha_{\mu} =43 (16)\times 10^{-10}$).
Although the situation regarding the $g_\mu -2$ 
has not been definitely settled, supersymmetry emerges as
a prominent candidate
in explaining the discrepancy between the Standard Model predictions and
experimental measurements, and in the sequel we concede
that this deviation accounted for SUSY. 
We find that this combination of the experimental information
from high energy physics and cosmology puts austere bounds on
 the parameter space of the  Constrained Minimal 
Supersmmetric Standard Model (CMSSM),
enabling us  to investigate the potential
of discovering SUSY, if it is based on CMSSM, at future  colliders
and direct DM search experiments.

\section{Neutralino relic density}
It has been argued that for large $\tan\beta$
the neutralino relic density ($\relic$) can be
compatible with the recent cosmological data which favour
small values for $\relic$. 
In this regime 
the neutralino ($\lsp$) pair annihilation 
through $s$-channel
pseudo-scalar Higgs boson ($A$) 
exchange leads to an enhanced annihilation cross sections
reducing significantly the relic 
density \cite{Drees}, 
while the heavy $CP$-even Higgs ($H$) exchange
is $P$-wave suppressed and not that important.
The importance of this mechanism, in conjunction with the recent
cosmological data which favour small values of the DM
relic density,
has been stressed in \cite{LNS,LNSd}. The same mechanism has been
also invoked  \cite{Ellis} where it 
has been shown that it enlarges the cosmologically
allowed regions. 
In fact cosmology does not put severe upper
bounds on sparticle masses, and soft masses can be in the TeV region,
pushing up the sparticle mass spectrum to regions that might escape detection
 in future planned accelerators. 
Such upper bounds are imposed, however, by
the recent $g_\mu -2$ E821 data \cite{E821,Narison} 
constraining the CMSSM in such a way that
supersymmetry will be accessible to LHC or other planned $e^{+}e^{-}$ linear
colliders if their center of mass energy is larger than 
about $1.2\TeV$ \cite{ENO}.
The bounds put by the $g_\mu -2$ has been 
the subject of intense phenomenological 
study the last few months \cite{ENO,g-2,Leszek,LS,LNSd2,Kneur}.

The $\lsp \lsp$ fusion to the pseudo-scalar Higgs boson, $A$, which
subsequently decays to a $b \bar{b}$ or a $\tau \bar{\tau}$, becomes the
dominant annihilation mechanism for large $\tan \beta$, when the pseudo-scalar
mass $m_A$ approaches twice the neutralino mass,  
$m_A \simeq 2 m_{\lsp}$.
In fact by increasing $\tan \beta$ the mass $m_A$ decreases, while the
neutralino mass remains almost constant, if the other parameters are kept
fixed. Thus  $m_A$ is expected eventually to enter into the regime in which
it is close to the pole value $m_A\,=\, 2 m_{\lsp}$, and the
pseudo-scalar  Higgs exchange dominates.
It is interesting to point out that in a previous analysis of the direct
DM searches \cite{LNSd}, we had stressed 
that  the contribution of the $CP$-even Higgs bosons exchange
to the LSP-nucleon scattering cross sections increases with $\tan \beta$.
Therefore in the large $\tan \beta$ 
regime one obtains the highest possible rates
for the direct DM searches. 
Similar results are presented in Ref.~\cite{Kim}. 
In the framework of the CMSSM
the chargino mass bound as well as the 
recent LEP Higgs mass bound \cite{LEP} already
exclude regions in which $\lsp$ has a large Higgsino component, and thus 
in the regions of interest the $\lsp$ is mainly a bino.  A bino is 
characterized by a very small coupling to the pseudo-scalar Higgs $A$, 
however the largeness of
$\tan \beta$ balances the smallness of its coupling giving a sizeable
effect when $m_A\;\simeq \; 2 m_{\lsp}$, making the $s$-channel
pseudo-scalar exchange mechanism important.

It becomes obvious from the previous discussion that an unambiguous and
reliable determination of the $A$-mass, $m_A$, 
is necessary in order to
to calculate the neutralino relic density especially 
in the large $\tan\beta$ region.
The  details of the 
procedure in calculating the spectrum of the CMSSM can be
found elsewhere \cite{LS,LNSd2}. Here 
we shall only briefly 
discuss some points which turn out to be
 essential for a correct determination of $m_A$.
In the constrained SUSY models,
such as the CMSSM, 
$m_A$ is not a free parameter but 
is determined once $m_0$, $M_{1/2}$, $A$ as
well as $\tan \beta$ and the sign of $\mu$  are given.
$m_A$  depends sensitively on the Higgs
mixing parameter, $m_3^2$, which is determined from minimizing 
the  one-loop corrected effective potential.
For large $\tan \beta$ the derivatives of the effective potential
with respect the Higgs fields, which enter into the minimization conditions, 
are plagued by terms which are large and hence potentially dangerous, making
the perturbative treatment untrustworthy.
In order to minimize the large $\tan \beta$
corrections we had better calculate the effective potential using as
reference scale the average stop scale
$Q_{\tilde t}\simeq\sqrt{m_{{\tilde t}_1} m_{{\tilde t}_2} }$ \cite{scale}. 
At this scale these terms are small and hence perturbatively valid.
Also for the calculation of the pseudo-scalar Higgs boson mass
 all the one-loop corrections must be taken into account. 
In particular, the inclusion of  those of the neutralinos and charginos  
yields  a result for $m_A$ that is scale independent and 
approximates the pole mass to better than $2 \%$  \cite{KLNS}.
A more significant correction, which 
drastically affects the pseudo-scalar mass 
arises from the gluino--sbottom and chargino--stop corrections to the bottom
quark Yukawa coupling  $h_b$ \cite{mbcor,wagner,BMPZ,arno}.
The proper resummation of these corrections
is important for a correct determination of $h_b$ \cite{eberl,car2},
and accordingly of the $m_A$.

In calculating the LSP
relic abundance, we solve the Boltzmann equation  
numerically using the machinery
outlined in Ref.~\cite{LNS}. In this calculation the coannihilation effects, 
in regions where $\tilde{\tau}_R$ approaches in mass the LSP, which is a high
purity Bino, are properly taken into account.

In what follows only the $\mu > 0$ case is considered. 
The $\mu<0$ case is not favored by the recent
$b \goes s  \gamma$ data, as well as by the observed discrepancy
of the $g_{\mu}-2$, if the latter is attributed to supersymmetry, and  
therefore we shall discard it.

\begin{figure}[t] 
\includegraphics[height=7.5cm,width=7.5cm]{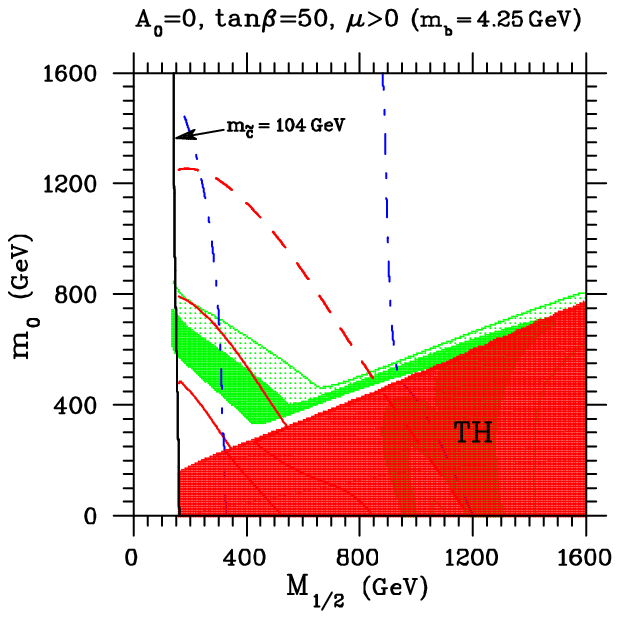}
\hspace*{.5cm}
\includegraphics[height=7.5cm,width=7.5cm]{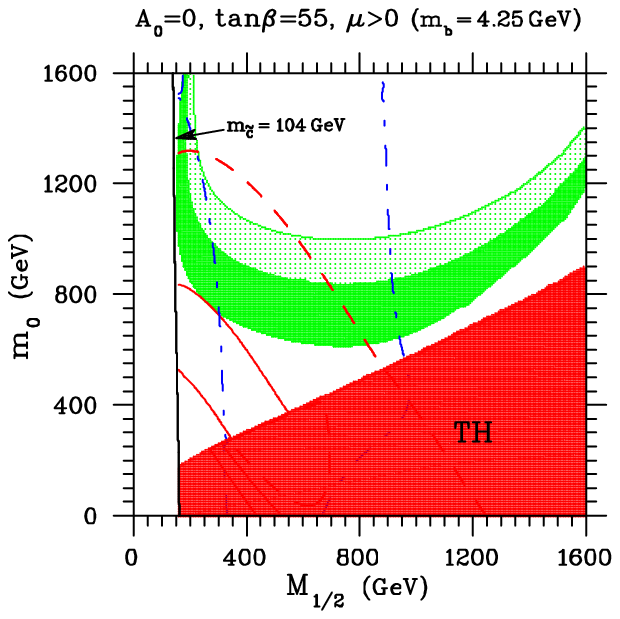}

\caption[]{
Cosmologically allowed regions of the relic density for two different values
 of $\tan \beta$ in the $(M_{1/2},m_0)$ plane. 
The remaining inputs are shown in each figure. 
The mass of the top is taken $175\GeV$. In the dark
green shaded area $0.08<\relic<0.18$. In the light green shaded
area $0.18<\relic<0.30\;$. The solid red lines mark the region within which
the supersymmetric contribution to the anomalous magnetic moment of the
muon is
$\alpha^{SUSY}_{\mu} = (43.0\pm 16.0) \times 10^{-10}$.
The dashed red line
is the boundary of the region for which the lower bound is moved to
$11\times 10^{-10}  < \alpha^{SUSY}_{\mu}$. 
The dashed-dotted blue lines are
the boundaries of the region $113.5 \GeV \leq m_{Higgs} \leq 117.0 \GeV$.}

\label{fig1}  
\end{figure}

In the panels shown in figure~\ref{fig1} we display our results by drawing 
the cosmologically
allowed region $0.08<\relic<0.18$ (dark green) in the $m_0, M_{1/2}$ plane 
for values of $\tan \beta$ equal to  $50$ and $55$ respectively.
Also drawn (light green) is the region $0.18<\relic<0.30$.
The default values for the masses of massive quarks are
$m_t = 175 \GeV, m_{\tau} = 1.777\GeV$ and $m_b(m_b) = 4.25\GeV$. 
The remaining inputs are shown on the top of each panel.
The solid red mark the region within which
the supersymmetric contribution to the anomalous magnetic moment of the
muon falls within the E821 range 
$ \alpha^{SUSY}_{\mu} = ( 43.0 \pm 16.0 ) \times 10^{-10}$.
The dashed red line
marks the boundary of the region when the more relaxed lower
bound $11 \times 10^{-10} \leq \alpha^{SUSY}_{\mu}$
is used, corresponding 
to the $2\sigma$ lower bound of the E821 range.
Along the blue dashed-dotted lines the light $CP$-even Higgs mass takes
values $113.5\GeV$ (left) and $117.0\GeV$ (right) respectively.
The line on the left
marks therefore the recent LEP bound on the Higgs mass \cite{LEP}.
Also shown is the chargino mass bound $104\GeV$
{\footnote {In the context of our analysis
focus point regions  \cite{focus} show up for
smaller values of the top mass.  
In any case the bulk of the focus point region 
appears  for rather large
values of $m_0$ and hence they
are not favoured by the $g_\mu - 2$ data.}}.
The shaded area (in red)
at the bottom of each figure, labelled by TH, is theoretically disallowed 
since the light stau is lighter than the lightest of the neutralinos.
From the displayed figures
we observe that for values of $\tan \beta$ up to $50$ the cosmological data 
put an upper bound on the parameter $m_0$. 
However, there is practically no such upper bound for the parameter $M_{1/2}$, 
due to the coannihilation effects \cite{Ellis} which allow for $M_{1/2} $
as large as $ 1700\GeV$ within 
the narrow coannihilation band lying above the theoretically disallowed
region.

For $\tan \beta = 55 $ a large region opens up within which the relic density
is cosmologically allowed. This is due to the pair annihilation of the
neutralinos through the pseudo-scalar Higgs exchange in the $s$-channel.
As explained before, for such high $\tan \beta$ the ratio $m_A / 2 m_{\lsp}$
approaches unity and the pseudo-scalar exchange dominates yielding
large cross sections and hence small neutralino relic densities. In this
case the lower bound put by the $g_\mu -2$ data 
cuts the cosmologically allowed
region which would otherwise allow for very large values of $m_0, M_{1/2}$. 
The importance of these corridors 
has been stressed in the analysis of \cite{Ellis}. 
However, in the analysis presented here  these show up at higher values
of the parameter $\tan \beta$.
We should remark at this point that in our analysis 
we use the value of $\alpha_{strong}(M_Z)$ as input and relax unification
of the $\alpha_3$ gauge coupling with the others. 
In the constrained scenario it is almost impossible to reconcile
gauge coupling unification with a value for $\alpha_{strong}(M_Z)$
consistent with experiment due to the low energy threshold effects. This
change affects drastically the values of other parameters and especially
that of the Higgsino ($\mu$) and  Higgs ($m_3^2$) mixing parameters that
in turn affect the pseudo-scalar Higgs boson mass which plays 
a dominant role.

For the $\tan \beta = 55$ case, 
close the highest possible value, and considering the conservative
lower bound on the muon's anomalous magnetic moment
$\alpha_{\mu}^{SUSY} \geq 11 \times 10^{-10}$
and values of
$\relic$ in the range $0.13\pm0.05$, 
we find that the point
with the highest value of $m_0$ is (in GeV) at
$(m_0, M_{1/2}) = (950, 300)$ and that with the highest value of 
$M_{1/2}$ is at $(m_0, M_{1/2}) = (600,750)$. 
The latter marks the 
lower end of
the line segment of the boundary $\alpha_{\mu}^{SUSY} = 11 \times 10^{-10}$ 
which amputates the cosmologically allowed stripe.
For the case displayed in the bottom right  panel of
the figure~\ref{fig1} 
the upper mass limits put on the LSP, and the lightest
of the charginos, stops and the staus are
$m_{\lsp} < 287, m_{{\tilde \chi}^{+}} <  539,  m_{\tilde t} < 1161,
m_{\tilde \tau} < 621$ (in $\GeV$). 
Allowing for $A_0 \neq 0$ values, the upper bounds put on $m_0, M_{1/2}$
increase a little and so do the aforementioned bounds on the sparticle
masses.
Thus it appears that the prospects of discovering CMSSM at  a
$e^{+} e^{-}$ collider with center of mass energy $\sqrt s = 800 \GeV$,
such as TESLA, are not guaranteed. 
However in the allowed regions 
the next to the lightest neutralino, ${\tilde{\chi}^{\prime}}$, has a mass
very close to the lightest of the charginos and hence the process
$e^{+} e^{-} \goes {\tilde{\chi}} {\tilde{\chi}^{\prime}}$,
with 
${\tilde{\chi}^{\prime}}$ subsequently decaying to
$ {\tilde{\chi}} + {l^{+}} {l^{-}}$ or 
$ {\tilde{\chi}}+\mathrm{2\,jets}$,  
is kinematically allowed for such large $\tan \beta$, provided
the energy is increased to at least $\sqrt{s} = 900 \GeV$. It should
be noted however that this channel proceeds via the $t$-channel exchange
of a selectron  is suppressed due to the heaviness of the exchanged
sfermion.

\begin{table}[t]
\begin{center}
\begin{tabular}{|c|c|c|c|c|c|} \hline \hline
$\tan\beta$ & $\lsp^0$ & $\tilde{\chi}^+$ & $\tilde{\tau}$ & $\tilde{t}$ & $h$ 
                                                                    \\ \hline
  10  &  108 (174) & 184 (306) & 132 (197) & 376 (686) &  115 (116) \\
  20  &  154 (255) & 268 (457) & 175 (274) & 603 (990) &  116 (118) \\
  30  &  191 (310) & 338 (560) & 212 (312) & 740 (1200) &  117 (118) \\
  40  &  201 (340) & 357 (617) & 274 (353) & 785 (1314) &  117 (119) \\
  50  &  208 (357) & 371 (646) & 440 (427) & 822 (1357) &  117 (119) \\
  55  &  146 (311) & 260 (563) & 424 (676) & 606 (1237) &  115 (117) \\ 
 \hline \hline
\end{tabular}
\end{center}

\vspace{.4cm}
\caption{Upper bounds, in GeV, 
on the masses of the lightest of the neutralinos,
charginos, staus, stops and  Higgs bosons for various values of
$\tan\beta$ if the the E821 bounds are imposed.
The values within brackets represent the same situation when the weaker
bounds
$11 \times 10^{-10}<\alpha_{\mu}^{SUSY}<75 \times 10^{-10}$
are used (see main text).}

\label{table1}
\end{table}

The situation changes, however, when the strict E821 limits are imposed 
$\alpha_{\mu}^{SUSY} = (43.0 \pm 16.0) \times 10^{-10}$. For instance 
in the $\tan \beta = 55$ case displayed in figure~\ref{fig1} there is no  
cosmologically allowed region which obeys this bound.
For the other cases, $\tan \beta < 50$, 
the maximum allowed $M_{1/2}$ is about $475\GeV$, occurring at
$m_0 \simeq 375\GeV$, and the maximum $m_0$ is $ 600\GeV$ when
$M_{1/2} \simeq 300\GeV$. The upper limits on the masses of the sparticles
quoted previously reduce to 
$m_{\lsp} < 192, m_{{\tilde \chi}^{+}}  <  353,  m_{\tilde t} < 775,
m_{\tilde \tau} < 436$ all in GeV.
However, these values 
refer to the limiting case $A_0 = 0$. Scanning the parameter space 
allowing also for $A_0 \neq 0$ we obtain the upper limits displayed in
the table~\ref{table1}. In this the unbracketed values correspond to the E821
limits on the $g_\mu -2$. 
For completeness we also display, within brackets, the
bounds obtained when the weaker lower bound
$\alpha_{\mu}^{SUSY} \geq 11 \times 10^{-10}$ is imposed.
We see that even at TESLA with
center of mass energy $\sqrt s = 800\GeV$, the prospects of discovering
CMSSM are guaranteed in the
$e^{+} e^{-} \goes {\tilde \chi}^{+} {{\tilde \chi}^{-}} $
if the E821 bounds are imposed.

\begin{figure}[t] 
\centering\includegraphics[height=8cm,width=9cm]{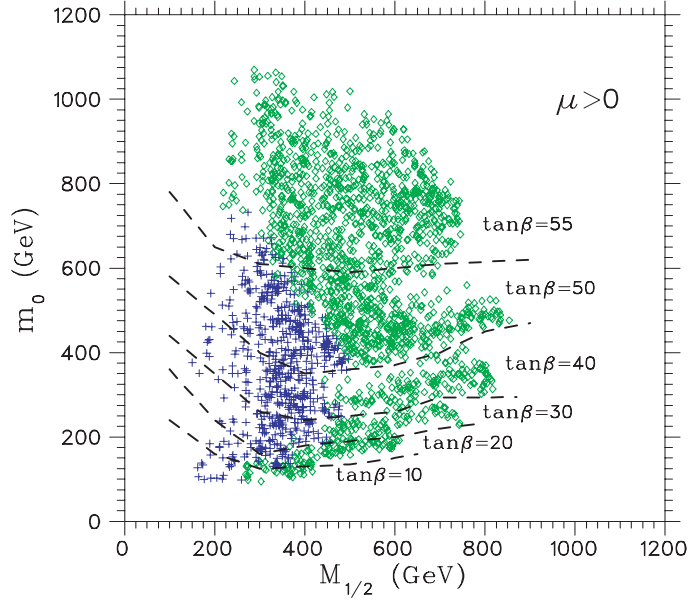}
\caption[]{
In  the $(M_{1/2},m_0)$ plane, we display all
points compatible with
$\alpha_{\mu}^{SUSY} = ( 43.0 \pm 16.0 ) \times 10^{-10}$ ($+$) and 
$11 \times 10^{-10}<\alpha_{\mu}^{SUSY}<75 \times 10^{-10}$
($\diamond$).
All the points are consistent with 
the cosmological bound $\relic = 0.13\pm 0.05$
and they  are grouped in regions, separated by dashed contours 
each of which is the boundary of $\tan \beta $ with the value shown
beneath. 
In the top region, designated by $\tan \beta = 55$,
the parameter $\tan \beta $ takes values between 50 and 55.
}
\label{fig2}  
\end{figure}  

In the figure~\ref{fig2} we display in the $(M_{1/2},m_0)$ plane
the points which are consistent both with the muon's anomalous magnetic
moment bounds mentioned before and cosmology, as well as with the other
accelerators data.
Each of the points is taken from a sample of 45,000 random points in the 
part of the parameter
space  defined by $m_0 < 1.5 \TeV$, $M_{1/2} < 1.5 \TeV$, 
$|A_0| < 1 \TeV$ and $2<\tan\beta<55$.
All the points are consistent with 
the cosmological bound $\relic = 0.13\pm 0.05$.
The  plus points (colored in blue) are those consistent with the E821 bound  
$ 27 \times 10^{-10} < \alpha_{\mu}^{SUSY} < 59 \times 10^{-10} $,
while the
diamonds (colored in green) are consistent with the  more relaxed  bound
$11 \times 10^{-10}< \alpha_{\mu}^{SUSY} <75 \times 10^{-10}$.
The points are grouped in regions, separated by dashed contours,
each of which constitutes  
the boundary of $\tan \beta$ with the value shown beneath. 
In the region designated as $\tan \beta = 55$ all points have
$55 > \tan \beta > 50$. 
It is seen clearly that only a few points in the $\tan \beta > 50$ case
can survive the E821 bound. For $\tan \beta < 50$  the parameter
$M_{1/2}$ cannot be larger than about $500\GeV$, attaining its maximum
value at $m_0 \simeq 400\GeV$ , and the maximum $m_0$ is $725 \GeV$ 
occurring at  $M_{1/2}\simeq 275\GeV$.
The upper limits put on $m_0, M_{1/2}$ 
result to the sparticle mass bounds displayed in the
table~\ref{table1}.

\begin{figure}[t] 

\centering\includegraphics[height=8cm,width=9cm]{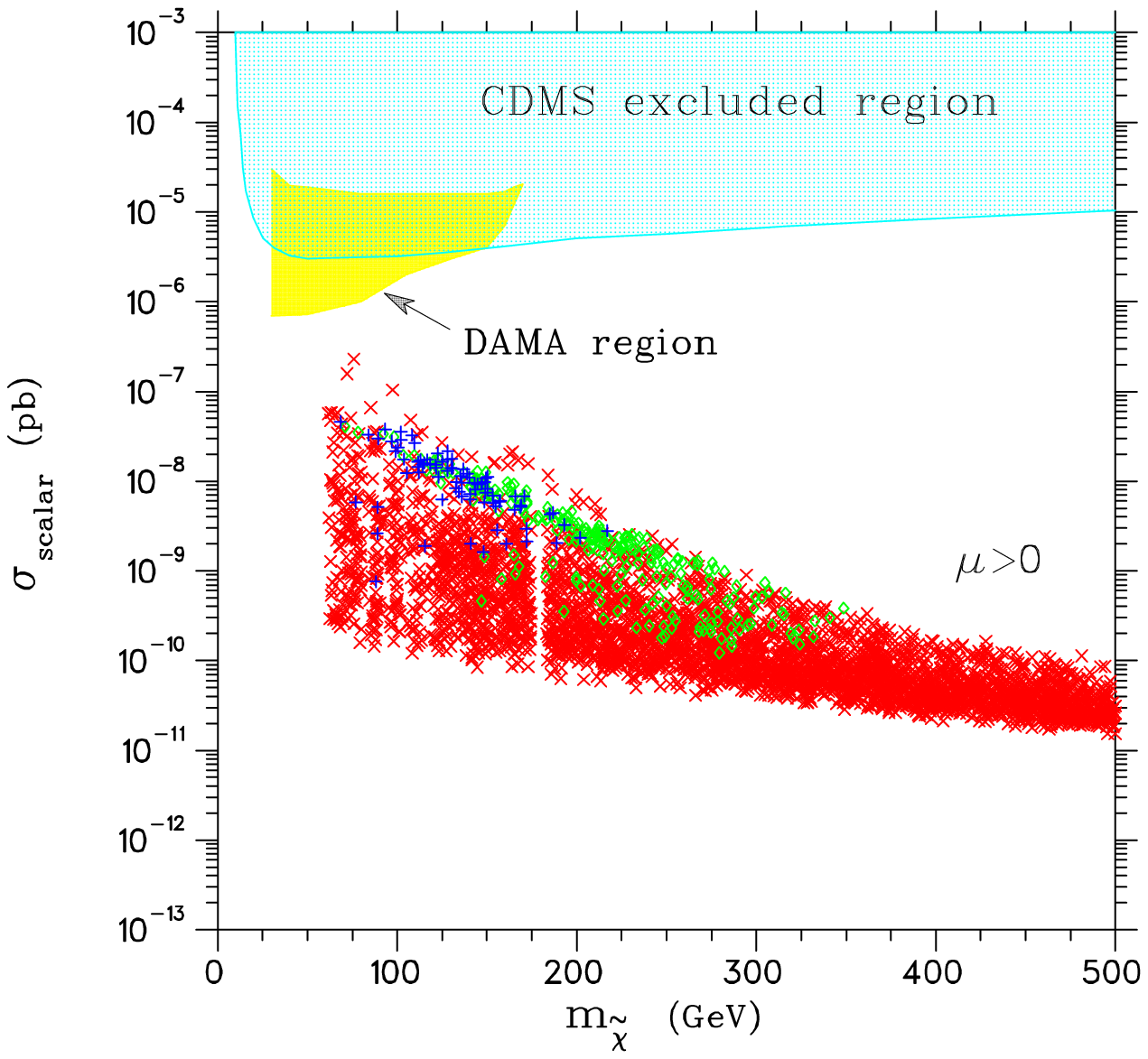}
\caption[]{
Scatter plot of
the scalar neutralino-nucleon cross section
 versus $\mlsp$, from a
random sample of 45,000 points.
On the top of the figure the CDMS excluded region and
the DAMA sensitivity region are illustrated.  
Pluses ($+$) (in blue colour) 
are points within the E821 experimental region
$\almuon = ( 43.0 \pm 16.0 ) \times 10^{-10}$ and
also cosmologically acceptable $\relic=0.13 \pm 0.05 $.
Diamonds ($\diamond$) (in green  colour) 
are also cosmologically
acceptable points, but with $\almuon$ within the
region $11 \times 10^{-10}<\almuon<75 \times 10^{-10}$.
Crosses ($\times$) (in red colour) represent the rest of the random sample.
The Higgs boson mass bound $m_h > 113.5 \GeV$ 
is properly taken 
into account.}

\label{fig3} 
\end{figure}  

\section{Direct Dark Matter searches}
We shall discuss now the impact of the $g_\mu -2$ measurements
and of the Higgs mass bound $m_h > 113.5 \GeV$
on the direct DM searches. We are using 
the same random sample 
as in figure~\ref{fig2} in order to
calculate the spin-independent,
$\lsp$-nucleon cross section ($\sxsec$).
In figure~\ref{fig3} we plot the scalar $\lsp$-nucleon   
 cross section as function of the LSP mass, $\mlsp$.
On the top of the figure the shaded region (in cyan colour) is
excluded by the CDMS experiment \cite{cdms}.
The DAMA sensitivity region (coloured in yellow) is
also plotted \cite{dama}. 
Pluses ($+$) (in blue colour) represent points 
which are both compatible
 with the E821 data $\almuon = (43.0\pm 16.0)\times 10^{-10}$
and the cosmological bounds for the neutralino relic density
$\relic = 0.13 \pm 0.05$. Diamonds ($\diamond$) (in green colour)
are points which are cosmologically acceptable with respect to
the aforesaid bounds, but the bound to the $\almuon$ has
been relaxed to its $2\sigma$ region, namely
 $11 < \almuon \times 10^{10} < 75 $.
The crosses ($\times$) (in red colour) represent the rest of the
points of our random sample.  Here the Higgs boson
mass bound, $m_h > 113.5 \GeV$ has been properly taken into account.
From this figure it is seen that the   
points which are compatible both the $g_{\mu}-2$ E821
and the cosmological data (crosses) yield cross sections
of the order of $10^{-8}-10^{-9}$ pb and  the maximum value of  the $\mlsp$
is about $200 \GeV$.
If one allows 
the $2 \sigma$ region of the $g_{\mu}-2$ bound the
lower bound of 
preferred cross sections is $10^{-10}$ pb and
correspondingly the upper bound of  $\mlsp$ is drifted up to $350\GeV$.

\begin{figure}[t] 

\centering\includegraphics[height=8cm,width=9cm]{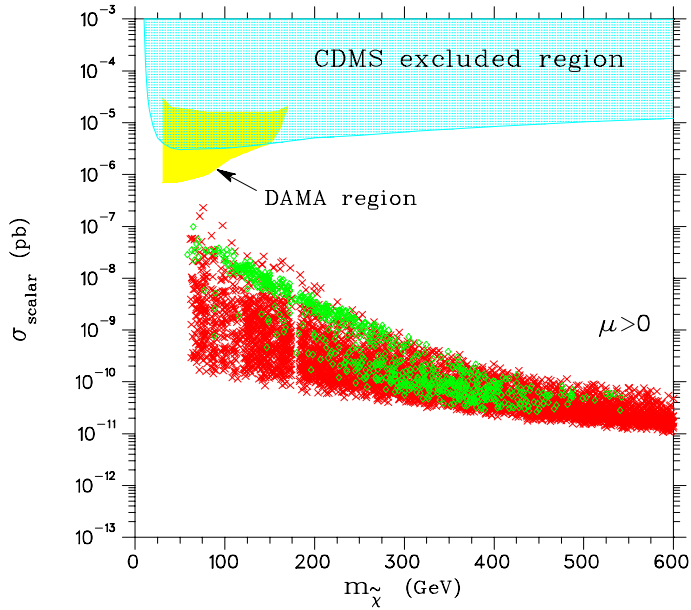}
\caption[]{
Scatter plot of
the scalar neutralino-nucleon cross section
 versus $\mlsp$, from a
random sample of Fig.~\ref{fig3}.  
Diamonds ($\diamond$) (in green colour) are  cosmologically
acceptable points, without putting an restriction from the
$\almuon$. Crosses ($\times$) (in red colour) represent points with
unacceptable $\relic$.}

\label{fig4} 

\end{figure}  

Comparing figure~\ref{fig3} and \ref{fig4} one can
realise how $g_{\mu}-2$ data constrain $\mlsp$ mass  to be up to $200 \GeV$ or
$350 \GeV$ for the $1\sigma$ or $2\sigma$ case respectively. 
In figure~\ref{fig4} we don't impose the constraints stemming
from $g_{\mu}-2$ data, therefore  due to the coannihilation processes
the cosmologically acceptable LSP mass can be heavier than
$500 \GeV$.  What is also important   to be noticed
about the direct searches of  DM is that imposing the $g_{\mu}-2$ data  
the lowest allowed $\lsp$-nucleon cross section increased by about
1 order of magnitude, from $10^{-11}$ pb to $10^{-10}$ pb.
Similar results are presented in Ref.~\cite{Keith}. 
This fact is very encouraging for the future
DM direct detection experiments, with sensitivities extending up to
$10^{-9}$ pb~\cite{Klapdor}.

\section{Conclusions} 
Concluding, we combined recent high 
energy physics experimental information,
like the anomalous magnetic moment of the muon measured at
 E821 Brookhaven experiment   and the light Higgs boson mass bound
from LEP, with the  cosmological data for DM.
By doing so we studied the imposed constraints on the
parameter space of the CMSSM and hence we assessed the potential
of discovering SUSY, if it is based on CMSSM, at future colliders
and DM direct searches experiments.
The bounds put on the sparticle spectrum can guarantee that in LHC  
but also in 
a $e^{+}e^{-}$ linear collider with center of mass energy
$\sqrt{s} = 800\GeV$, such as TESLA, CMSSM can be discovered.
The guarantee for a linear collider  with this energy is
lost in a charged sparticle final state channel, 
if the lower bound on the value of $g_\mu -2 $ is lowered to its
$\approx 2 \sigma$ value, but not for the LHC. 
In this case only by increasing the
center of mass energy to be $\simeq 1.2 \TeV$, a
$e^{+}e^{-}$ linear collider can find CMSSM in 
$\tilde{\tau} \, {\tilde{\tau}}^*$ or
${\tilde \chi}^{+} {{\tilde \chi}^{-}} $ channels.
The impact of the E821 experiment's result along with the
bound on Higgs mass is also significant 
for the direct DM searches.
We found that the maximum value of the spin-independent $\lsp$-nucleon
cross section attained is of the order of $10^{-8}$ pb.
Moreover this cross section can not be lower than $10^{-10}$ pb,
which is very promising for the forthcoming direct DM
experiments.

\Acknowledgements
A.B.L. acknowledges support from HPRN-CT-2000-00148
and HPRN-CT-2000-00149 programmes. He also thanks the University of Athens
Research Committee for partially supporting this work. 
D.V.N. acknowledges support by D.O.E. grant DE-FG03-95-ER-40917.
V.C.S. acknowledges support  by a Marie Curie Fellowship of the EU
programme IHP under contract HPMFCT-2000-00675.

\end{document}